\documentclass{article}

\usepackage[preprint]{neurips_2024}

\usepackage[utf8]{inputenc}
\usepackage[T1]{fontenc}
\usepackage{hyperref}
\usepackage{url}
\usepackage{booktabs}
\usepackage{amsmath}
\usepackage{amssymb}
\usepackage{graphicx}
\usepackage{xcolor}
\usepackage{natbib}
\graphicspath{{figures/}}
\usepackage{newunicodechar}
\newunicodechar{×}{\ensuremath{\times}}
\newunicodechar{−}{\ensuremath{-}}
\newunicodechar{Δ}{\ensuremath{\Delta}}
\newunicodechar{≥}{\ensuremath{\geq}}
\newunicodechar{—}{\textemdash}
\newunicodechar{–}{\textendash}
\newcommand{\rhonote}{Corrected in version~2: this correlation carries no
information beyond the two instruments' shared dependence on error rate. See
the erratum in Section~\ref{sec:audit}.}

\hypersetup{colorlinks=true, linkcolor=blue!50!black,
            citecolor=blue!50!black, urlcolor=blue!50!black}

\title{Bistable by Construction: Wall-Clock-Calibrated State Monitors\\
Have No Moment-Detection Regime at Agent Cadence}

\author{%
  Manvendra Modgil\\
  Modint Intelligence\\
  \texttt{manvendramodgil.ai@gmail.com}\\
}
\date{June 2026}

\begin{document}
\maketitle

\begin{abstract}
Runtime monitors for autonomous agents commonly threshold an accumulated
internal state---a behavioural baseline, a drift statistic, or, in our prior
work, a modelled affective state. We previously reported a State Saturation
Trap: threshold-on-state triggers over a continuous affect engine become
near-constant alarms on SWE-bench debugging agents~\cite{P1}. A post-release
audit of our replay pipeline found that the engine received $\Delta t = 0$
between actions, so its exponential decay never operated: the published trap is
a pure-accumulator result. We correct the record (erratum in \cite{P1}-v2) and
treat the flaw as an experiment design. The key variable it exposes is one the
monitoring literature does not track: whether the monitor's dynamics are
calibrated in \emph{sample time} (per observation, as in classical CUSUM) or in
\emph{wall-clock time} (half-lives in seconds, as in affect models and EMA
baselines). On fixed-rate streams these coincide; on agent streams---where
inter-action time varies by orders of magnitude across deployments---they do
not. A pre-registered sweep over uniform inter-action intervals
($\Delta t \in \{0..600\}$\,s) on 20 trajectories shows the
wall-clock-calibrated level trigger has exactly two regimes: at
$\Delta t \le 1$\,s it is a constant alarm (trap holds 20/20; median 18 firings
per run); at $\Delta t \ge 60$\,s it is silent (the state never reaches
threshold). Every trajectory's critical $\Delta t$ lies in $(1, 30]$\,s.
Instrumented runs of a scriptable single-tool-call agent loop (gpt-5.4-mini)
measure inter-action latency at median 1.53\,s (p90 2.33\,s)---the measured
cadence sits firmly inside the trap regime, vindicating \cite{P1}'s empirical
finding under a corrected
mechanism---and replaying real ordered latency sequences shows the regimes are
stable within a run (19/20 clean single-crossing accumulators; bounded flicker
only when bursts are scaled into the critical band, exploratory). The regime is
therefore selected by the deployment's latency profile, not by within-run
noise. The structure is a property of the calibration class, not the engine: a
minimal wall-clock leaky accumulator over the raw per-action error stream
reproduces the same cliff (20/20 trap at $\Delta t = 0$; 0/20 crossings at
$\Delta t \ge 60$; critical band of the same order of magnitude, Spearman
$\rho = 0.530$ with the engine's)\footnote{\rhonote}, while a sample-time CUSUM over the identical
stream is exactly invariant across the entire $\Delta t$ grid (20/20 identical
fire patterns). A rising-edge trigger with hysteresis at the identical
threshold fires 0--3 times per trajectory in every condition tested, including
a live instrumented demo. A zero-parameter balance model,
$r \ge (1 - 2^{-\Delta t/T_{1/2}})(\theta - B)$, captures the transition's order
of magnitude but not per-trajectory variation. We conclude that
wall-clock-calibrated leaky-integrator monitors admit no operating regime in
which they act as moment detectors on agent streams; transition detection
escapes the trap at every cadence, though---consistent with \cite{P1}'s
inter-rater results---it does not thereby recover human intervention timing.
\end{abstract}

\section{Introduction}
\label{sec:intro}

A growing family of runtime oversight systems decides when to interrupt an
autonomous agent by thresholding an accumulating internal quantity: a predicted
risk score, a behavioural baseline, a drift statistic, or a modelled affective
state. Our prior work~\cite{P1} used a continuous 18-dimensional affect engine
as a diagnostic probe over SWE-bench-Verified debugging trajectories and
reported that threshold-on-state triggers degenerate into near-constant
alarms---the State Saturation Trap---alongside a second finding that the
supervised target itself (human intervention timing) is a low-reliability
construct.

The present paper's organising observation is a calibration distinction the
monitoring literature does not track. Classical sequential detectors define
their dynamics in \emph{sample time}: CUSUM's drift parameter is subtracted per
observation, so the statistic is indifferent to how much wall-clock time
separates observations. Affect models, physiological analogies, and EMA-style
behavioural baselines define their dynamics in \emph{wall-clock time}:
half-lives in seconds, calibrated to human or process timescales. On a
fixed-rate sensor stream the two calibrations are equivalent up to a constant.
On an agent action stream they are not: inter-action time is seconds in a fast
tool loop, tens of seconds under heavy test suites or large prefills, and
minutes under human approval gates or long reasoning pauses---one to two orders
of magnitude of variation \emph{across deployments} of the same agent. We show
that this variation silently selects the qualitative behaviour of any
wall-clock-calibrated leaky-integrator monitor.

Concretely, this paper does five things. (1) It corrects~\cite{P1}: an audit
found the engine's decay term never operated in our replays
(Section~\ref{sec:audit}), so the published trap describes a pure accumulator.
(2) It converts the flaw into the central experiment: a pre-registered
counterfactual sweep over inter-action time, revealing the level trigger is
bistable---constant alarm or dead, transition within one order of magnitude of
$\Delta t$ (Section~\ref{sec:sweep}). (3) It measures where real agents sit, via
an instrumented agent loop, and replays real ordered latency sequences to
show the regimes are stable within a run
(Sections~\ref{sec:cadence}--\ref{sec:burst}). (4) It demonstrates the
structure is a property of the calibration class, not the engine: a minimal
wall-clock accumulator over the raw error stream reproduces the cliff while a
sample-time CUSUM over the identical stream is exactly $\Delta t$-invariant
(Section~\ref{sec:class}). (5) It evaluates pre-registered transition-aware
triggers: edge detection escapes the trap in every condition including live
operation, but no trigger recovers human-annotated intervention timing,
consistent with~\cite{P1} (Section~\ref{sec:transition}).

The contribution is a mechanism-level account of why wall-clock-calibrated
threshold monitors fail on agent streams, with a falsifiable scaling rule, a
measured cadence anchor, and a class-level demonstration---rather than a new
detector benchmark.

\section{Setup}
\label{sec:setup}

We reuse the engine, observer, and trigger stack of~\cite{P1} entirely
unmodified: an 18-dimensional affect vector with per-emotion exponential decay
toward baseline $B = 0.10$ (frustration half-life $T_{1/2} = 150$\,s),
momentum-modulated decay, conflict rebalancing, and an energy cap; an observer
mapping each agent action (thought, tool call, observation) to event deltas via
eight fixed heuristic rules; and the A6 absolute-threshold triggers, of which
\texttt{sustained\_frustration} (fire iff frustration $\ge 0.7$) is the focus.
All thresholds remain at their~\cite{P1} values; no constant anywhere in the
stack was changed for this study. Hypotheses, trigger definitions, instrument
constants, and the $\Delta t$ grid were written to pre-registration documents
before the corresponding results were generated; ordering is evidenced by
filesystem modification times and by the verbatim agreement between each
pre-registered hypothesis and the wording scored in the results reports
(Appendix~\ref{app:prereg}). The artefact tree was not under version control
during the study and is placed under git only for this release, so the ordering
rests on those two checks rather than on signed commit timestamps. The two
exploratory items added later are labelled as such where they appear
(Sections~\ref{sec:burst}--\ref{sec:class}).

\textbf{The instrument class.} We use ``wall-clock-calibrated leaky integrator''
for any monitor statistic of the form $s' = \mathrm{decay}(s, \Delta t) +
\mathrm{step}(\text{event})$, where decay is parameterised in seconds (e.g.,
exponential with a fixed half-life) and the statistic is sampled once per agent
action. HEART is one member; Section~\ref{sec:class} constructs a minimal second
member and a sample-time control outside the class.

\textbf{Data.} 20 trajectories from the public 20250514\_aime\_coder
SWE-bench-Verified submission: the 5 pilot trajectories of~\cite{P1} plus 15
selected by a fixed a-priori rule (first 15 by instance id with 25--70 actions
that parse cleanly), spanning astropy and django instances, 25--59 actions each.
A distribution check shows the original pilot was broadly representative but
skewed long (median 44 vs.\ 32 actions). Separately, 5 live instrumented runs
(Section~\ref{sec:cadence}) provide wall-clock timing.

\textbf{Signal sparsity.} Between 25\% and 79\% of actions per trajectory
produce no event delta at all (median 52\%); the observer's rules do not match
most edits and shell commands. On those actions the state evolves by decay
alone---a fact whose significance the next section establishes.

\section{The $\Delta t = 0$ Audit and Erratum}
\label{sec:audit}

While instrumenting the pipeline for this study we audited the time step the
engine receives between actions. The trajectory parser assigns each action a
synthetic ordinal timestamp, but it is never read on the replay path: the
adapter applies events through a code path whose decay tick receives
$\Delta t = 0$ and returns immediately. Decay therefore never operated in any
replay reported in~\cite{P1}. We verified this empirically: on the trajectory
with 78.6\% zero-signal actions, all 43 zero-signal actions leave the 18-vector
unchanged to machine precision, including a 10-action stretch at full saturation
that decay should have visibly relaxed.

Consequences: every~\cite{P1} firing rate and replay-consistency number is
unaffected (they are deterministic functions of the $\Delta t = 0$ pipeline),
but the mechanistic account---agents accumulate negative affect faster than it
decays---was wrong. There was no race between accumulation and decay; there was
only accumulation. The published trap is the $\Delta t = 0$ limit of a family of
systems indexed by inter-action time. We issue an erratum in~\cite{P1}-v2 and
devote the rest of this paper to mapping that family.

A decay-vs-event decomposition over the persisted state streams attributes
100.0\% of state variation to events on all trajectories---by construction, not
as a model property---and shows the engine is not a pass-through of rule outputs
(mean $|\text{raw} - \text{realised}|$ delta of 0.044--0.093 per event-bearing
action from momentum, conflict rebalancing, and normalisation).

\paragraph{Erratum to this paper (v2).} Version~1 of this paper contains three
errors, corrected in this version. Each correction is quoted below from the
errata record of~\cite{P3}, which refers to this paper as Paper~2.

\emph{(1) How the cadence was measured.} Version~1 described the timing runs of
Section~\ref{sec:cadence} as hook-instrumented (abstract;
Sections~\ref{sec:intro} and~\ref{sec:cadence}).
\begin{quote}
The five runs were not logged by the Claude Code hook. They were driven by a scriptable loop issuing one tool call per turn to gpt-5.4-mini, which wrote the same PostToolUse-schema record natively after each observation. The hook, which produces the identical format, was installed but was not the source of these timestamps. Paper 2 does not name the model. No number changes: the timestamp is taken at the same point in each cycle, after the tool's observation.
\end{quote}
This version names the model. The correction bears only on how far the
measured cadence generalises, which Section~\ref{sec:limitations} already
bounded (``a lower bound for frontier loops---the deployments most likely to
enter the transition band are exactly the ones not measured''). The abstract,
Sections~\ref{sec:intro}, \ref{sec:cadence} and~\ref{sec:limitations}, and the
Instrumentation appendix are corrected in this version.

\emph{(2) The cross-instrument correlation.} Version~1 reports that
per-trajectory critical intervals correlate between the two instruments at
Spearman $\rho = 0.530$ (abstract; Section~\ref{sec:class}).
\begin{quote}
This correlation carries no information beyond the two instruments' shared dependence on error rate. It equals the product of each instrument's correlation with error rate to three decimals (0.899 × 0.593 = 0.533), and the partial correlation controlling for error rate is −0.007 (permutation p = 0.97). Paper 2's class-level conclusion is unaffected: it rests on both instruments showing the same two regimes on identical streams (20/20 in the alarm regime at Δt = 0; none crossing threshold at Δt ≥ 150 s), not on this correlation.
\end{quote}
Both passages are footnoted in this version.

\emph{(3) The scope of the abstract's cadence claim.} Version~1's abstract said
that real autonomous coding cadence sits firmly inside the trap regime.
\begin{quote}
This holds for the harness, model and session lengths measured, and Paper 2's own limitations section said so: the cadence came from `a small-model loop on small repos (a lower bound for frontier loops—the deployments most likely to enter the transition band are exactly the ones not measured)'. The abstract claimed more than that limitation allowed. A companion paper finds that a production agent harness driving frontier models acts more slowly than Paper 2's loop, in a comparison that also changes model vendor and task set, and that on its shorter sessions the monitor was starved of evidence rather than trapped: no session reached the clamp. Whether the trap closes depends on cadence and session length; it is not a general property of autonomous coding.
\end{quote}
This version's abstract narrows the claim to the measured cadence.

\section{The Uniform-Cadence Sweep}
\label{sec:sweep}

We replay all 20 trajectories with a synthetic uniform inter-action interval
$\Delta t$ injected before each action (engine untouched; decay invoked
explicitly from the replay caller), over the pre-registered grid
$\Delta t \in \{0, 1, 5, 15, 30, 60, 150, 300, 600\}$\,s. We measure: whether
frustration ever crosses 0.7; persistence (the fraction of post-first-crossing
actions still $\ge 0.7$); and fire counts for the level trigger (A6
\texttt{sustained\_frustration}) and the edge trigger (T3,
Section~\ref{sec:transition}).

\begin{table}[t]
\centering
\caption{Trap persistence vs $\Delta t$ (20 trajectories).}
\label{tab:persistence}
\begin{tabular}{rccc}
\toprule
$\Delta t$ (s) & crossing 0.7 & mean persistence (crossers) & \# trap holds ($\ge 90\%$) \\
\midrule
0   & 20/20 & 100.0\% & 20 \\
1   & 20/20 & 100.0\% & 20 \\
5   & 19/20 & 93.4\%  & 17 \\
15  & 14/20 & 63.4\%  & 2  \\
30  & 5/20  & 15.6\%  & 0  \\
60  & 1/20  & 2.6\%   & 0  \\
$\ge 150$ & 0/20 & --- & 0 \\
\bottomrule
\end{tabular}
\end{table}

\begin{table}[t]
\centering
\caption{Level vs edge trigger, fire counts (min/median/max over 20
trajectories).}
\label{tab:levelvedge}
\begin{tabular}{rcc}
\toprule
$\Delta t$ (s) & A6 level & T3 edge (net) \\
\midrule
0   & 5 / 18 / 47 & 1 / 1 / 1 \\
1   & 3 / 18 / 47 & 1 / 1 / 1 \\
5   & 0 / 16 / 47 & 0 / 1 / 2 \\
15  & 0 / 4 / 28  & 0 / 1 / 2 \\
30  & 0 / 0 / 5   & 0 / 0 / 1 \\
60  & 0 / 0 / 1   & 0 / 0 / 1 \\
$\ge 150$ & 0 / 0 / 0 & 0 / 0 / 0 \\
\bottomrule
\end{tabular}
\end{table}

The level trigger has two regimes and nothing else. At $\Delta t \le 1$\,s it is
the~\cite{P1} constant alarm. At $\Delta t \ge 60$\,s the modelled state never
reaches threshold and the trigger is silent---not because the agent recovered,
but because human-calibrated decay ($T_{1/2} = 150$\,s) drains a trickle of
event input (mean realised frustration input 0.015--0.036 per action) faster
than it arrives. Per-trajectory critical $\Delta t$ (persistence $< 50\%$) lies
in $(1, 5]$ for 3 trajectories, $(5, 15]$ for 8, and $(15, 30]$ for 9---the
entire population transitions within one order of magnitude, and none survives
past 30\,s. There is no $\Delta t$ at which A6 fires a small number of times at
moments: its median fire count goes $18 \to 16 \to 4 \to 0$ across
$\Delta t = 1 \to 5 \to 15 \to 30$, passing through ``fires occasionally'' only
as a narrow transition between regimes. A leaky integrator at fast cadence is an
accumulator; at slow cadence it is empty.

\begin{figure}[t]
\centering
\includegraphics[width=\textwidth]{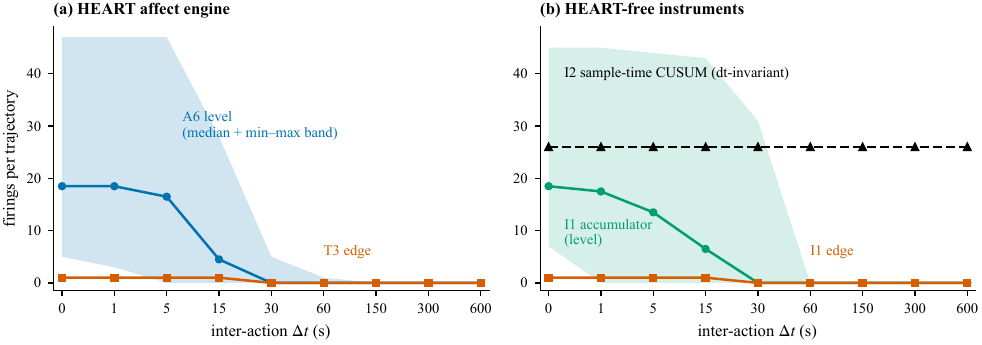}
\caption{The thesis in one figure, three monitor archetypes on the same input.
\textbf{(a)} HEART: the A6 level trigger's median fire count (blue) with min--max
band collapses from a constant alarm to silence across the $\Delta t$ grid,
while the T3 edge trigger (orange) is flat at $\le 1$. \textbf{(b)} HEART-free
instruments over the raw error stream: the I1 wall-clock accumulator (green)
shows the same cliff, its edge variant (orange) stays $\le 2$, and the I2
sample-time CUSUM (black, dashed) is exactly $\Delta t$-invariant at 26 fires.
The failure mode lives in the calibration choice, not the state model.
$x$-axis is the discrete $\Delta t$ grid (linear index, labelled in seconds).}
\label{fig:fig1}
\end{figure}

\begin{figure}[t]
\centering
\includegraphics[width=0.62\textwidth]{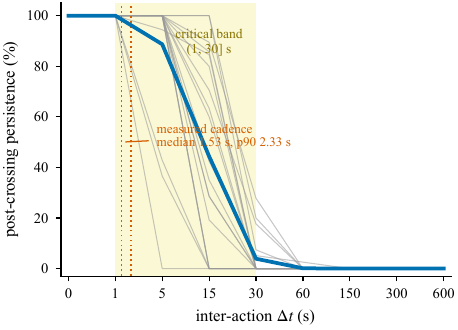}
\caption{Post-crossing persistence vs $\Delta t$: 20 thin per-trajectory lines
(grey), mean (blue), with the critical band $(1, 30]$\,s shaded. Measured real
agent cadence (median 1.53\,s, p90 2.33\,s; dotted) sits deep inside the
constant-alarm regime.}
\label{fig:fig2}
\end{figure}

\section{Measured Real Cadence}
\label{sec:cadence}

The sweep's $\Delta t$ is synthetic, and the SWE-bench traces record no
wall-clock times (we verified the submission's public artefacts carry none). We
therefore instrumented an agent loop directly: a scriptable loop issues one tool
call per turn to gpt-5.4-mini and, after each tool's observation is returned,
appends an ISO wall-clock timestamp, tool name, and result status to a local
log. The record uses the same schema, taken at the same point in the cycle, as a
Claude Code \texttt{PostToolUse} hook released alongside it; the hook was not
the source of these timestamps (version~1 said otherwise; see the erratum in
Section~\ref{sec:audit}). Five real debugging runs on cloned open-source
Python repositories (verified bug injection, real test suites; 7--32 actions per
run) yield a pooled $n = 65$ inter-action sample: \textbf{median 1.53\,s, p90
2.33\,s, max 15.87\,s} (the maximum is a full slow test-suite step; test
execution dominates the tail at p90 $= 15.5$\,s by tool type, while reads and
edits sit at 1--2\,s). Only 3.1\% of actions exceed 5\,s; none exceeds 30\,s.
Lag-1 autocorrelation of $\Delta t$ is approximately zero (0.033): heavy steps
did not cluster. An earlier toy probe and token-based reconstructions of the
original traces (medians 1.1--4.4\,s under assumed decode rates) corroborate the
range from independent directions.

Placed on Table~\ref{tab:persistence}'s grid, measured real cadence sits at the
$\Delta t \in [1, 5]$\,s grid points---\textbf{inside the constant-alarm regime
for every trajectory}, with rare transient excursions toward the band's edge
during slow-suite steps. Two consequences. First,~\cite{P1}'s empirical finding
is vindicated: the $\Delta t = 0$ replays approximated the regime real fast
agents actually occupy; the mechanism was wrong, the regime was right. Second,
the dead regime ($\Delta t \ge 60$\,s) is not reached by this workload---but it
is plausibly reached by deployment patterns we did not measure: long
reasoning-model thinking time, CI pipelines, container builds, and human
approval gates routinely exceed 60\,s. We state this as an extrapolation, not a
measurement. The selecting variable is the \textbf{deployment latency profile}:
the same monitor on the same agent behaviour is a constant alarm in a fast tool
loop and silent in a human-gated loop. A monitor designer who does not log
inter-action latency cannot know which monitor they have deployed.

\textbf{Measurement limitation, stated plainly:} the instrumented loop used a
small, fast model (gpt-5.4-mini) on small repositories; frontier-model loops with heavy prefill
and long reasoning are slower, so our median is a lower bound for the
deployments of greatest interest---which strengthens the deployment-profile
concern rather than weakening it.

\section{Real Burst Structure: Regimes Are Stable Within a Run}
\label{sec:burst}

Uniform $\Delta t$ leaves open whether realistic non-uniform latency makes the
monitor \emph{flicker}---crossing and re-crossing threshold as heavy steps drain
the state---which would make its firings a latency detector wearing an affect
costume. We tested this with pre-registered conditions per trajectory: C1,
constant $\Delta t$ at the measured median (1.53\,s); C2, i.i.d.\ lognormal
$\Delta t$ fit to the measured distribution (10 seeds); C3, the five real
ordered latency sequences from Section~\ref{sec:cadence}, tiled to trajectory
length with burst structure intact (420 replays total). Pre-registered
hypotheses: H5, the level trigger flickers (median crossing count $> 2$) on
trajectories whose critical interval brackets the median; H6, T3 stays $\le 3$
net fires in all conditions.

\textbf{Disclosure of design timing.} C1--C3 and H5/H6 were committed before
execution; their parameters necessarily derive from Section~\ref{sec:cadence}'s
measurements, as pre-specified. An analytic pre-check (decay from the
frustration clamp to below 0.7 requires ${\sim}88$\,s of silence; the largest
measured gap is 15.9\,s) predicted C1--C3 would not reach the flicker regime.
Rather than over-read the expected null, we added one exploratory
condition---C4, the same real burst sequences time-scaled into each trajectory's
critical band---specified after the pre-check but before any replay was run, and
labelled exploratory throughout. We verified, against the committed
pre-registration document, that the H5/H6 wording and metrics are unchanged from
the version committed before execution, and that only C4 and the
analytic-expectation note were added afterward (see Appendix~\ref{app:prereg}
for the provenance check and its caveat).

\textbf{Results.} H5 is not supported: under C2 and C3, all eligible
trajectories show median crossing counts $\le 1$. Under real ordered bursts
(C3), 19/20 trajectories are clean single-crossing accumulators; the single
exception (a trajectory hovering near 0.7 exactly when the 16\,s heavy step
lands) double-crosses in 3 of 5 sequences---a boundary case, not flicker. Per
the pre-registration's falsification clause, the knife-edge framing is
accordingly softened: within a run, at measured cadence, the regime is stable.
The exploratory C4 supplies the other half: with identical burst \emph{shapes}
shifted into the critical band, 9/20 trajectories multi-cross (up to 4
crossings; flicker index up to 3)---the in-band sensitivity is real, but it is
bounded re-crossing, never wild oscillation. H6 is supported, and beyond its
scope: T3 stays at $\le 1$ net fire in every C1--C3 cell and $\le 3$ even in C4.

Together, Sections~\ref{sec:sweep}--\ref{sec:burst} replace ``latency noise
selects the failure mode'' with the more precise claim the data supports:
\textbf{the regimes are bistable across deployments and stable within them.}
Latency variation inside a fast loop does not toggle the monitor; moving the
deployment's latency profile across the $(1, 30]$\,s band does.

\begin{figure}[t]
\centering
\includegraphics[width=0.62\textwidth]{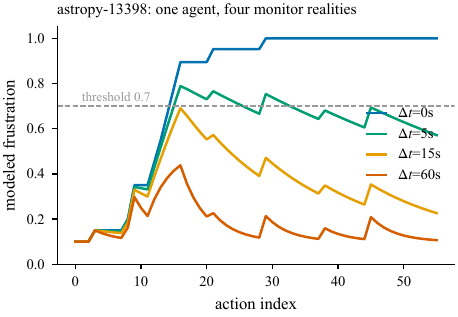}
\caption{Frustration vs action index for one trajectory
(\texttt{astropy-13398}) at $\Delta t \in \{0, 5, 15, 60\}$\,s: the same agent
behaviour, four monitor realities, from a permanent alarm to a state that never
reaches threshold.}
\label{fig:fig3}
\end{figure}

\section{The Class, Not the Engine}
\label{sec:class}

A natural objection: perhaps the bistability is an artefact of HEART---its 18
dimensions, its heuristic rules, its conflict mechanics. We test the class-level
claim with two minimal instruments over a HEART-independent input: the raw
per-action binary error indicator $e_i$ (verified identical across all
$\Delta t$-variant replays), on the same 20 trajectories and the same
$\Delta t$ grid, with all constants pre-registered.

\textbf{I1, a wall-clock leaky accumulator:}
$s_i = \mathrm{clamp}_{01}(s_{i-1} e^{-\lambda \Delta t} + 0.15\, e_i)$, with
the same half-life as HEART frustration ($\lambda = \ln 2 / 150$) by design,
isolating the calibration choice; level trigger at 0.7, edge trigger with
hysteresis at 0.7/0.5. \textbf{I2, a sample-time CUSUM:}
$g_i = \max(0, g_{i-1} + e_i - 0.10)$, fire at $g \ge 1.0$---by construction
$\Delta t$-free.

\textbf{Results.} I1 reproduces the two-regime structure: the trap holds on
20/20 trajectories at $\Delta t = 0$ (18/19 crossers at $\Delta t = 1$), and the
state never reaches threshold on 20/20 at $\Delta t \ge 60$---the same cliff, on
a one-dimensional accumulator with no emotions, no rules, and no engine. I2 is
exactly invariant: identical fire-index sets at all nine $\Delta t$ values on all
20 trajectories (26 fires throughout, the flat reference line of
Figure~\ref{fig:fig1}). I1's edge trigger fires at most 2 times anywhere. One
pre-registered conjunct failed and is reported as such: we predicted I1's
critical intervals would fall in the same $(1, 30]$ band as HEART's; 16/20 do,
while 4 land in adjacent bins ($(30, 60]$ for three high-error-rate
trajectories, $(0, 1]$ for the sparsest). Per-trajectory critical $\Delta t$s
correlate between instruments at Spearman $\rho = 0.530$.\footnote{\rhonote} The honest statement:
\textbf{cadence bistability is a property of wall-clock-calibrated leaky
integrators sampled at agent cadence}---the band's exact placement depends on
input stream and step size, within the same order of magnitude. The
error-stream statistics (rates 0.09--0.64) and a step size larger than HEART's
typical realised deltas account for the placement differences.

The three-archetype contrast is the paper in one figure: the wall-clock level
trigger's cliff, the edge trigger's flat $\le 2$ line, and sample-time CUSUM's
exact invariance, all on the identical input stream. The failure mode lives in
the calibration choice, not in any particular state model.

\section{A First-Order Scaling Rule (and Its Limits)}
\label{sec:scaling}

Balancing per-action decay drain against mean per-action event input $r$ at
threshold $\theta$ gives the trap condition
$r \ge (1 - 2^{-\Delta t/T_{1/2}})(\theta - B)$, hence a predicted critical
$\Delta t = -(T_{1/2}/\ln 2)\ln(1 - r/(\theta - B))$ with zero free parameters.
Computed from each trajectory's realised input statistics, predictions land at
5.6--13.4\,s---the right order of magnitude, inside the observed population
band---but only 8/20 fall within their trajectory's interval-censored
observation. An exploratory correction (estimating $r$ from the pre-saturation
region only, motivated by clamp censoring of 0--67\% of input; specified after
seeing the original residuals and labelled as such) repairs the diagnosed
mechanism yet \emph{reduces} interval accuracy to 6/20, trading
under-predictions for over-predictions. We take this as informative failure: a
mean-rate model carries the correct dimensionless structure---input rate
$\times$ half-life vs.\ threshold elevation, the quantity any leaky-integrator
monitor designer should compute---but per-trajectory transitions are governed by
input burstiness that a mean rate cannot express. We deliberately stop here
rather than fit burst-aware parameters post hoc.

\begin{figure}[t]
\centering
\includegraphics[width=0.62\textwidth]{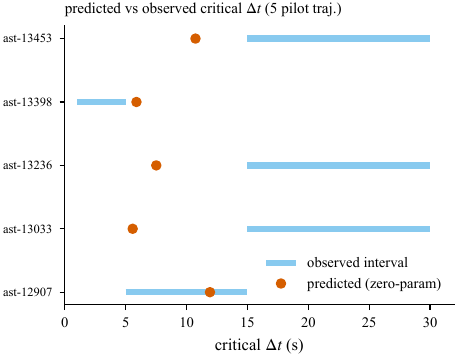}
\caption{Zero-parameter predicted critical $\Delta t$ (orange points) against
interval-censored observations (blue bars) for the 5 pilot trajectories: right
order of magnitude, but only a minority land inside their interval.}
\label{fig:fig4}
\end{figure}

\section{Transition Triggers Escape the Trap but Not the Timing Problem}
\label{sec:transition}

We pre-registered four transition-aware triggers over the same frozen state
stream: T1 velocity (windowed first difference $\ge 0.5$ on the 5-emotion
negative sum), T2 acceleration (second difference $\ge 0.3$), T3 rising-edge
saturation entry (fires on upward crossing of the identical 0.7 frustration
threshold; re-arms below 0.5), T4 plateau-no-recovery, all behind a 5-action
refractory wrapper, thresholds fixed before any result existed.

\textbf{Saturation escape (pre-registered H1): partially confirmed, cleanly for
T3.} Across the $n = 20$ set, the full uniform grid, all non-uniform conditions
including exploratory C4, and the HEART-free instrument of
Section~\ref{sec:class}, edge detection fires 0--3 times per trajectory, at
saturation onset (re-fires only when decay legitimately pulls the state below
the hysteresis floor and it re-crosses). At $\Delta t = 0$, where A6 fires on
18--82\% of actions, T3 fires once. Since T3 and A6 share the exact threshold
value, the comparison isolates level- vs.\ edge-detection: the entire pathology
of A6 is the \emph{level} primitive, not the threshold, the emotion model, or
the data. T1 mostly escapes (post-saturation fire rate 2--27\%); T2 never fires
anywhere---its pre-registered second-difference threshold is mis-scaled to the
engine's per-event delta magnitudes (0.05--0.2), a calibration lesson we report
rather than repair; T4 degenerates (26--88\% of post-saturation actions),
confirming that plateau detection at saturation reproduces the level trigger's
failure with extra steps.

\begin{figure}[t]
\centering
\includegraphics[width=0.7\textwidth]{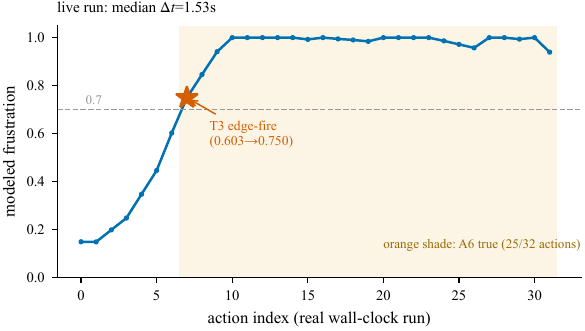}
\caption{Live instrumented run (real wall-clock $\Delta t$, median 1.53\,s):
modelled frustration ramps from 0.10 to saturation; T3 fires once at the rising
edge (star, $0.603 \to 0.750$) while A6 is true on 25 of 32 actions (orange
shading)---the level-vs-edge contrast in live operation.}
\label{fig:fig5}
\end{figure}

\textbf{Live operation.} The instrumented loop of Section~\ref{sec:cadence} also
ran the frozen engine and T3 online with real elapsed $\Delta t$. On the
32-action run, T3 fired exactly once, at the rising edge (frustration
$0.603 \to 0.750$), while A6 was true on 25 of 32 actions---the level-vs-edge
contrast in a single live log (Figure~\ref{fig:fig5}).

\textbf{Timing alignment (pre-registered H2): failed, as the pre-registration's
falsification clause anticipated.} On the one human-labelled trajectory, the
transition triggers' net firings overlap the $\ge 2$-annotator consensus set at
2/9, and both hits come from the degenerate T4 surviving an arbitrary refractory
subsample---a stopped-clock effect we decline to count as alignment. Given
\cite{P1}'s central finding that three trained annotators agree on intervention
points only marginally above chance (location $\alpha = +0.047$), this is the
coherent outcome: transition triggers repair the \emph{detector}; nothing can
repair an unreliable \emph{target}. Solving saturation and solving timing are
different problems, and only the first is solved here.

\section{Related Work}
\label{sec:related}

\textbf{Runtime oversight of agents.}
A growing line of work interrupts an agent by checking its trajectory against
safety conditions. AgentSpec~\cite{agentspec} compiles user-specified rules
(trigger--predicate--enforce) and fires when a predicate matches the current
action; AgentHarm~\cite{agentharm} benchmarks harmful agent behaviour that such
monitors aim to catch. These systems threshold per-action predicates rather
than an accumulated, time-decayed statistic, and so lie outside the
leaky-integrator class our result concerns. ProbGuard~\cite{probguard} is the
closest in spirit---it predicts the probability of reaching an unsafe state and
intervenes when that probability crosses a threshold---but the quantity it
thresholds is an \emph{instantaneous} predicted probability recomputed from the
current symbolic state at each step, not a statistic that integrates past
evidence with a wall-clock decay. It is therefore also outside our class: it
has no half-life to mis-calibrate against cadence, and our bistability result
does not apply to it. Our prior work~\cite{P1} is the wall-clock-calibrated
member whose failure this paper dissects.

\textbf{Sequential change detection.}
The classical home of edge-versus-level detection is the cumulative-sum chart
of Page~\cite{page1954}, with asymptotic optimality established by
Lorden~\cite{lorden1971} and the modern landscape surveyed by Truong et
al.~\cite{truong2020}. CUSUM is defined in \emph{sample time}---its drift term
is subtracted once per observation---which is exactly why our Section~7 control
(I2) is cadence-invariant: it is classical CUSUM behaving classically. We claim
no novelty for edge detection or for CUSUM; the contribution is the
\emph{calibration-mismatch} analysis that separates sample-time from
wall-clock-time dynamics at \emph{agent} cadence, where inter-action intervals
vary across deployments by orders of magnitude. The two-threshold hysteresis
our T3 trigger uses is the Schmitt trigger~\cite{schmitt1938}, the canonical
bistable detector with separate set/reset levels.

\textbf{Calibrated-decay monitors and label variation.}
Wall-clock-calibrated leaky integrators---exponential-moving-average behavioural
baselines and drift statistics with fixed half-lives---are common in production
agent and service monitoring; they are precisely the family our analysis places
at risk when the sampling cadence is not the cadence the half-life was tuned
for. Finally, our evaluation of whether any trigger \emph{recovers human
intervention timing} inherits the framing of LLM-as-judge
evaluation~\cite{zheng2023} and of human label
variation~\cite{plank2022,krippendorff2004}: as in~\cite{P1}, the target itself
is a low-reliability construct, so repairing the detector (Section~9) does not,
and cannot, repair the target.

\section{Limitations}
\label{sec:limitations}

Synthetic uniform $\Delta t$ in Section~\ref{sec:sweep} (bounded by
Sections~\ref{sec:cadence}--\ref{sec:burst}'s measured and replayed real timing,
but the SWE-bench traces' own timing remains unrecoverable); measured cadence
from a small-model loop (gpt-5.4-mini) on small repos (a lower bound for frontier loops---the
deployments most likely to enter the transition band are exactly the ones not
measured); the dead-regime claim for slow deployments is an extrapolation from
the sweep, not a measurement; one harness format and one model family's traces;
the class-level demonstration uses one additional instrument over one input
stream (binary errors)---broader instrument and stream coverage would strengthen
it, and the band's placement is instrument-dependent (one pre-registered band
conjunct failed, reported in Section~\ref{sec:class}); timing-alignment
evaluation rests on one labelled trajectory pending expanded annotation; the
scaling rule is order-of-magnitude only; T2/T4 show that pre-registration
without pilot calibration risks degenerate cells, a cost of the no-tuning
discipline we accept; C4 is exploratory and its in-band flicker result awaits
confirmatory replication.

\section{Conclusion}
\label{sec:conclusion}

A wall-clock-calibrated leaky-integrator monitor has no cadence regime in which
it detects moments on agent streams: below the critical band it is a constant
alarm, and above it the monitor is dead. The regimes are stable within a run, and the
transition band sits where real deployments live---so the deployment's latency
profile, a quantity monitor designers neither control nor typically log,
silently selects which failure is deployed. The structure is a property of the
calibration class: a one-dimensional accumulator over raw errors shows the same
cliff, while a sample-time CUSUM over the identical stream is exactly
cadence-invariant. Measured real agent cadence (median 1.53\,s) sits firmly in
the constant-alarm regime, vindicating the empirical content of our earlier
report under a corrected mechanism. The correct primitive at any cadence is
transition detection---a rising edge with hysteresis fired 0--3 times per
trajectory in every condition tested, including live operation---but detecting
transitions is not the same as timing interventions, which remains a
low-reliability human construct. We arrived here by auditing our own published
pipeline, finding its decay term inert, and pre-registering the counterfactuals
the flaw made possible; we commend the genre.

\bibliographystyle{plainnat}
\bibliography{references}

\appendix
\section{Appendix}

\subsection{Trajectory selection}
15 selected, 11 skipped for length, 0 parse failures (first 15 by instance id
with 25--70 actions that parse cleanly, excluding the 5 pilot trajectories).
Full skip log released with the artefacts.

\subsection{Per-trajectory tables}
Full saturation table ($\Delta t = 0$), per-trajectory critical-$\Delta t$
intervals, and zero-signal coverage are released as
\texttt{SCALE\_REPORT.md} and the \texttt{fig\_data/} CSVs.

\subsection{Pre-registration provenance and hypothesis scorecard}
\label{app:prereg}
Pre-registration documents for all phases are released verbatim
(\texttt{PREREGISTRATION\_\allowbreak PHASE7.md}, \texttt{PREREGISTRATION\_\allowbreak PHASE8.md}).
Full hypothesis scorecard: H1 partial (T3 yes; T1 borderline; T2 vacuous; T4
no), H2 failed, H3 superseded by the $\Delta t$ audit, H4 pending annotation, H5
not supported (with the C4 exploratory complement), H6 supported, H7 partial
(two-regime supported; strict band conjunct failed 16/20), H8 supported (exact
invariance 20/20), H9 supported (max 2).

\paragraph{Provenance check for the Section~\ref{sec:burst} disclosure.} The released
\texttt{PREREGISTRATION\_PHASE7.md} states H5 and H6 verbatim as ``\emph{H5:
under C2/C3, A6 exhibits flicker (median crossing count $>2$) on trajectories
whose uniform-$\Delta t$ critical interval brackets the $\Delta t$ median}'' and
``\emph{H6: T3 net fire count remains $\le 3$ under all conditions},'' which are
exactly the hypotheses scored in \texttt{NONUNIFORM\_REPORT.md}; C4 and the
``analytic expectation'' subsection are present as a separately labelled
\emph{exploratory} block. The artefact tree was \emph{not} under version control
during the study, so no git diff of this file exists from the study period; the
repository is placed under git only for this release. The evidence supporting
``written before execution'' is therefore (i) the file's text matches the
report's scored wording exactly, and (ii) filesystem modification times order
the pre-registration (\texttt{2026-06-10 21:26}) before the results report
(\texttt{2026-06-10 21:30}). We rely on these two checks rather than on signed
commit timestamps, which are stronger provenance but were not available during
the study; Section~\ref{sec:setup} is worded accordingly.

\subsection{Scaling rule}
Derivation and both estimators (full-trajectory $r$ and pre-saturation $r$),
with censoring diagnostics, in \texttt{ANALYTIC\_DT\_REPORT.md}.

\subsection{$\Delta t$ audit}
Code-path trace with line references and the erratum text in
\texttt{DT\_AUDIT.md}.

\subsection{Instrumentation}
The five timing runs were logged natively by the scriptable loop
(\texttt{heart\_instrument/agent\_runner.py}, gpt-5.4-mini, one tool call per
turn), which records the model's text with each action. The released Claude Code
\texttt{PostToolUse} hook writes the same record format for readers' own
sessions; its design, the tool-name mapping table, and its converter
limitations (empty thoughts, because hooks do not expose reasoning text; A9
reasoning-feature triggers therefore inapplicable to hook traces) apply to hook
traces, not to the five timing runs. Timing distributions per run and per tool
type in \texttt{data/live\_runs/}.

The loop writes a record after every tool observation whether or not the call
failed, so the five timing runs are complete. The released hook, as first
distributed, was not: Claude Code fires \texttt{PostToolUse} only after a tool
call succeeds and \texttt{PostToolUseFailure} after one fails, so a logger
registered for \texttt{PostToolUse} alone never records a failed command.
Readers instrumenting their own sessions should register it for both events;
the repository now does so. No number in this paper is affected.

\subsection{Non-uniform replay}
Condition construction, seeds, per-condition tables, and the boundary case in
\texttt{NONUNIFORM\_REPORT.md}.

\subsection{Generality instruments}
Error-stream statistics per trajectory, I1/I2 definitions and unit tests, full
Tables A--D, and the Spearman computation in \texttt{GENERALITY\_REPORT.md}.

\end{document}